\documentclass[12pt,indent]{article}
\usepackage{a4,latexsym,amsmath}
\usepackage[latin1]{inputenc}
\usepackage{float}
\usepackage{natbib}
\usepackage{graphics}
\usepackage[english]{babel}
\usepackage{tabularx}
\usepackage{lscape}
\usepackage{multirow}
\usepackage{rotating}
\usepackage{dsfont}
\usepackage{subfig}
\usepackage[table]{xcolor}
\usepackage{%
  longtable,
  array,
  booktabs,
  dcolumn,
  rotating,
  shortvrb,
  tabularx,
  units,
  multirow
}
\usepackage{calc}
\usepackage{ifthen}
\newenvironment{tabularsmall}
{ \footnotesize \sffamily \tabular } {
\endtabular
\normalfont }

\newcommand{\E}{\operatorname{E}}      
\newcommand{\var}{\operatorname{var}}

\newcommand{\logit}{\operatorname{logit}}

\newcommand{\NBV}{\operatorname{NB}}

\newcommand{\alphab}{\boldsymbol{\alpha}}

\newcommand{\betab}{{\boldsymbol{\beta}}}

\newcommand{\gammab}{\boldsymbol{\gamma}}
\newcommand{\pib}{{\boldsymbol{\pi}}}

\newcommand{\thetab}{\boldsymbol{\theta}}

\newcommand{\xb}{\boldsymbol{x}}

\newcommand{\Yb}{\boldsymbol{Y}}

\newcommand{\blanco}[1]{}

\def\d{\displaystyle}

\begin{document}
\sloppy

\makeatletter
\renewcommand{\section}{\@startsection{section}{1}{\z@}%
        {-3.5ex \@plus -1ex \@minus -.2ex}%
        {1.5ex \@plus.2ex}%
        {\reset@font\Large\sffamily}}
\renewcommand{\subsection}{\@startsection{subsection}{1}{\z@}%
        {-3.25ex \@plus -1ex \@minus -.2ex}%
        {1.1ex \@plus.2ex}%
        {\reset@font\large\sffamily\flushleft}}
\renewcommand{\subsubsection}{\@startsection{subsubsection}{1}{\z@}%
        {-3.25ex \@plus -1ex \@minus -.2ex}%
        {1.1ex \@plus.2ex}%
        {\reset@font\normalsize\sffamily\flushleft}}
\makeatother



\newsavebox{\tempbox}
\newlength{\linelength}
\setlength{\linelength}{\linewidth-10mm} \makeatletter
\renewcommand{\@makecaption}[2]
{
  \renewcommand{\baselinestretch}{1.1} \normalsize\small
  \vspace{5mm}
  \sbox{\tempbox}{#1: #2}
  \ifthenelse{\lengthtest{\wd\tempbox>\linelength}}
  {\noindent\hspace*{4mm}\parbox{\linewidth-10mm}{\sc#1: \sl#2\par}}
  {\begin{center}\sc#1: \sl#2\par\end{center}}
}



\def\R{\mathchoice{ \hbox{${\rm I}\!{\rm R}$} }
                   { \hbox{${\rm I}\!{\rm R}$} }
                   { \hbox{$ \scriptstyle  {\rm I}\!{\rm R}$} }
                   { \hbox{$ \scriptscriptstyle  {\rm I}\!{\rm R}$} }  }

\def\N{\mathchoice{ \hbox{${\rm I}\!{\rm N}$} }
                   { \hbox{${\rm I}\!{\rm N}$} }
                   { \hbox{$ \scriptstyle  {\rm I}\!{\rm N}$} }
                   { \hbox{$ \scriptscriptstyle  {\rm I}\!{\rm N}$} }  }

\def\d{\displaystyle}

\title{Transition Models for Count Data: a Flexible Alternative to Fixed Distribution Models}
\author{Moritz Berger$^*$ and Gerhard Tutz$^{**}$ \vspace{0.5cm} \\
\small{ $^*$Institut f\"{u}r Medizinische Biometrie, Informatik und Epidemiologie,} \\ \small{Universit\"{a}tsklinikum Bonn, Sigmund-Freud-Stra{\ss}e 25, 53127 Bonn, Germany} \vspace{0.1cm} \\  
{\small $^{**}$ Ludwig-Maximilians-Universit\"{a}t M\"{u}nchen,} \\ \small{Akademiestra{\ss}e 1, D-80799 M\"{u}nchen, Germany}} 
 
\maketitle

\begin{abstract} 
\noindent
A flexible semiparametric class of models is introduced that offers an alternative to classical regression models for count data as the Poisson and negative binomial model, as well as to more general models accounting for excess zeros that are also based on fixed distributional assumptions. The model allows that the data itself determine  the distribution of the response variable, but, in its basic form, uses a parametric term that specifies the effect of explanatory variables. In addition, an extended version is considered, in which the effects of covariates are specified nonparametrically. The  proposed model and traditional models are compared by utilizing several real data applications.
\end{abstract}

\section{Introduction}\label{introduction}
In many applications the  response variable of interest is a nonnegative integer or count which one wants to relate  to a set of covariates. Classical regression models are the Poisson model and the negative binomial model, which can be embedded into the framework of generalized linear models \citep{McCNel:89}. More general models use the generalized Poisson distribution, the double Poisson distribution or the Conway-Maxwell-Poisson distribution \citep{Consul:1998, zou2013evaluating, sellers2010flexible}. Specific models designed to account for excess zeros are the hurdle model and the zero-inflation model. Concise overviews of modeling strategies were given by \citet{KleZei:2008}, \citet{hilbe2011negative}, \citet{cameron2013regression} and \citet{hilbe2014modeling}. Specific models using distributions beyond the classical ones were considered, among others,  by  \citet{JoeZhu:2005}, \citet{GscCza:2006}, \citet{nikoloulopoulos2008modeling} and \citet{rigby2008framework}. More recently, existing count regression approaches were compared, for example, by \citet{hayat2014understanding}, \citet{payne2017approaches} and \citet{maxwell2018modelling}.

Most of the established models assume that a fixed distribution holds for the response variable conditional on the values of covariates, and the mean (and possibly the variance parameter) is linked to a linear function of the covariates. Various methods have been proposed to estimate the parametric effect of covariates on the counts \citep{cameron2013regression}.  
Specifying a distribution of the counts, however, can be rather restrictive, and the validity of inference tools depends on the correct specification of the distribution. 

To address this issue, we propose a class of models that does not require   specifying a fixed distribution for the counts. Instead, the form of the distribution is determined by parameters that reflect the tendency to higher counts, which has the effect that the fitted distribution is fully data-driven. For the estimation of the  parameters that determine the distribution penalized maximum likelihood estimation procedures are proposed. The proposed models automatically account for zero inflation, which typically calls for more complex models, see, for example, \citet{loeys2012analysis}. Interpretation of the parameters is kept simple, since the conceptualization uses that counts typically result from a process, where the final count is the result of increasing numbers. 

The effect of covariates on the response is modeled  by a linear term. This enables an easy interpretation of the regression coefficients in terms of multiplicative increases or decreases of the counts. The proposed models are semiparametric in nature, because the distribution of the response is modeled in a flexible way adapted to the data while the effect of covariates is modeled parametrically. The models are also extended to allow for smoothly varying coefficients in a nonparametric fashion. 

A main advantage of the proposed model class is that it can be embedded into the framework of binary regression. This implies that standard software for maximum likelihood estimation of generalized additive models \citep{Wood:2006c} can be used for model fitting.  

The rest of the article is organized as follows: In Section \ref{sec:count} classical models for count data are briefly reviewed. In Section \ref{sec:transition} the model class is introduced and penalized maximum likelihood estimation methods are described. Section~\ref{sec:app} is devoted to illustrative applications. In Section \ref{sec:tran} the model is extended to allow for more flexible effects of covariates, which are not necessarily restricted to linear effects. Details on implementation and software are given in Section \ref{sec:software}. Section \ref{sec:conc} summarizes the main findings of the article. 

\section{Classical Models for Count Data}\label{sec:count}

Let 
$Y_i\in\{0,1,2,\dots\}$ denote the response variable and $\xb_i^T=(x_{i1},\hdots,x_{ip})$ a vector of covariates of an i.i.d. sample with $n$ observations. In generalized linear models (GLMs) one specifies a distributional assumption for $Y_i|\xb_i$ and a structural assumption that links the mean $\mu_i=\E(Y_i|\xb_i)$ to the covariates. The structural
assumption in GLMs has the form
\[
\mu_i=h(\xb_i^T\betab) \quad \text{or} \quad g(\mu_i)=\xb_i^T\betab\,,
\]
where $\betab=(\beta_1,\hdots,\beta_p)^\top$ is a set of real-valued coefficients, $g$ is a known link function and $h=g^{-1}$ denotes the response function, see, for example, \citet{McCNel:89}, \citet{FahTut:2001}.

\subsection{Poisson and Negative Binomial Model}
Popular models for count data are the Poisson model and the negative binomial model. 
The Poisson model assumes that $Y_i|\xb_{i}$ follows a Poisson distribution~$P(\mu_i)$, where the mean and the variance are given by $\mu_i$, respectively. The most widely used model uses the canonical
link function by specifying
\begin{align}\label{eq:poisson}
\mu_i=\exp(\xb_i^T\betab) \quad \text{or} \quad \log(\mu_i)=\xb_i^T\betab\,.
\end{align}
In model \eqref{eq:poisson} the expressions $\exp(\beta_1),\hdots,\exp(\beta_p)$ have a simple interpretation in terms of multiplicative increases or decreases of $\mu_i$.  
More generally on can assume that $Y_i|\xb_i$ follows a negative binomial
distribution $\NBV(\nu,\mu_i)$ with probability density function (p.d.f.)
\[
  P(Y_i|\xb_i) =
    \frac{\Gamma(Y_i+\nu)}{\Gamma(\nu)\Gamma(Y_i+1)}\left(\frac{\mu_i}{\mu_i+\nu}\right)^{y_i} \left(\frac{\nu}{\mu_i+\nu}\right)^{\nu}\,,
\]
with mean and variance given by
\begin{equation}\label{eq:NB}
\mu_i=\exp(\xb_i^T\betab) \quad \text{and} \quad \var(Y_i|\xb_i)=\mu_i+\mu_i^2/\nu\,.
\end{equation}
While the mean is the same as for the simple Poisson model, the
variance exceeds the Poisson variances by $\mu_i^2/\nu$, which may
be seen as a limiting case $(\nu\rightarrow\infty)$.


\subsection{Zero-Inflation Model}

In some applications one observes more zero counts than is
consistent with the Poisson  or negative binomial model;
then data display overdispersion through excess zeros. This happens in cases in which the population consists of  two subpopulations, the non-responders who are "never at risk" with counts $Y_i=0$ and the responders who are at risk with counts $Y_i \in \{0,1,\dots\}$. 
Formally, the zero-inflated density function is a mixture of distributions. With $C_i$ denoting the
class indicator of subpopulations ($C_i=1$ for responders and
$C_i=0$ for non-responders) one obtains the mixture distribution
\[
P(Y_i|\xb_i)=P(Y_i|C_i=0)\,\pi_i+P(Y_i|\xb_i,C_i=1)\,(1-\pi_i)\,,
\]
where $\pi_i=P(C_i=0)$ are the mixing probabilities. Typically one assumes that a classical count data model, for example the Poisson model \eqref{eq:poisson}, holds 
for the responders, that is, one assumes $Y_i|\xb_i,C_i=1 \sim P(\mu_i)$, and a binary model, for example the logistic model, determines class membership. 
Then the link between responses and covariates is determined by the two predictors
\begin{align} \label{eq:Zero}
\log(\mu_i)=\xb_i^T\betab \quad \text{and} \quad \logit(\pi_i)=\xb_i^T\gammab\,,
\end{align}
where $\gammab=(\gamma_1,\hdots,\gamma_p)^\top$ is a second set of real-valued coefficients. 

More general models, in which the Poisson distribution is replaced by the generalized Poisson distribution have been considered by 
\citet{FamSin:2003}, \citet{FamSin:2006}, \citet{Guptaetal:2004}, \citet{Czadoetal:2007} and \citet{CzaMin:2009}.  Estimation procedures for zero-inflated  models are available in the R package~\textbf{pscl}~\citep{Zeietal:2008}.

\subsection{Hurdle Model}

An alternative model that is able to account for excess zeros is the
hurdle model~\citep{Mullahy:86, CreLoo:90}. The model specifies two processes that generate the zero counts and the positive counts. It combines a truncated-at-zero count model (left-truncated at $Y_i=1$) which is employed for positive counts and a binary model or censored count model (right-censored at $Y_i=1$) which determines whether the response is zero or positive, i.e., if the ``hurdle is crossed''. Formally, the hurdle model is
determined by
\begin{align*} 
P(Y_i|\xb_i)=\begin{cases}f_1(0|\xb_i)&\quad \text{if} \; Y_i=0\,,\\f_2(Y_i|\xb_i)\frac{1-f_1(0|\xb_i)}{1-f_2(0|\xb_i)}&\quad \text{if} \; Y_i>0\,,\end{cases}
\end{align*}
where $f_1$ determines the binary decision between zero and a positive outcome.  If the hurdle is crossed, the response is determined by the truncated count model with p.d.f.
\[
P(Y_i|\xb_i)=f_2(Y_i|\xb_i)/(1-f_2(0|\xb_i))\,, \qquad Y_i=1,2,\hdots\,.
\]
If $f_1=f_2$, the model collapses to the so-called \textit{parent process} $f_2$. The model is quite flexible, because it allows for both under- and overdispersion. 
For example, in the hurdle Poisson model where both $f_1$ and $f_2$ correspond to
Poisson distributions with means $\mu_1$ and $\mu_2$, the link between responses and covariates is determined by the two predictors 
\begin{align} \label{eq:hurdle} 
\mu_{i1}=\exp(\xb_i^T\betab) \quad \text{and} \quad \mu_{i2}=\exp(\xb_i^T\gammab).
\end{align}
The Poisson and geometric hurdle model have been examined by
\citet{Mullahy:86}, negative binomial hurdle models have been considered by \citet{PohUlr:95}. Estimation procedures for hurdle models are available in the R package \textbf{pscl}~\citep{Zeietal:2008}.

\section{The Transition Model for Count Data} \label{sec:transition}

In particular the Poisson and the negative binomial model have a
simple structure with a clearly defined link between the covariates
and the mean. The models however are rather restrictive since they
assume that the distribution of $Y_i|\xb_i$ is known and fixed. A
strict parametric form is assumed for the whole support~$\{0,1,2,\dots\}$ and typically just the dependence of the mean on the predictors is specified by the model.

\subsection{The Basic Transition Model}

The approach proposed here is much more flexible and does not assume
a fixed distribution for the response. It focusses on the modelling
of the transition between counts. In its simplest form, it assumes
for $Y_i \in \{0,1,2,\dots\}$
\begin{equation}\label{eq:basmodel}
P(Y_i>r|Y_i\ge r,\xb_i)= F(\theta_r+\xb_i^T\betab)\,, \quad r=0,1,\hdots\,,
\end{equation}
where $F(\cdot)$ is a fixed distribution function. The parameters $\theta_r$ represent intercept coefficients and $\betab^T=(\beta_1,\hdots,\beta_p)$ is a vector of regression coefficients. One models the transition probability
$\delta_{ir}=P(Y_i>r|Y_i\ge r,\xb_i)$, which is the conditional probability that the number of counts is larger than $r$ (i.e., transition to a higher value than $r$) given the number of counts is at least $r$. These probabilities are determined by a classical binary
regression model. For example, if $F(\cdot)$ is the logistic
distribution function, one uses the binary logit model. 

In general, a distribution  of count data $Y_i$ can be characterized by the probabilities on the support $\pi_{i0},\pi_{i1}, \hdots\,$, where $\pi_{ir}=P(Y_i=r)$, \textit{or} the (conditional) transition probabilities $\delta_{i0},\delta_{i1}, \hdots\,$ given by
\[
\delta_{ir}=P(Y_i>r|Y_i\ge r)= \frac{1-\pi_{i0}-\hdots-\pi_{ir}}{1-\pi_{i0}-\hdots-\pi_{i,r-1}}\,.
\]
The transition model (\ref{eq:basmodel})  specifies the transition probabilities.  If no covariates are present, any discrete distribution with support $\{0,1,2,\hdots\}$ can be represented by the model, determined by the intercept parameters $\theta_0, \theta_1, \hdots$. In the
presence of covariates the intercepts represent the basic
distribution of the counts, which is modified by the values of the
covariates. Thus, the functional form of the count distribution is not restricted.
In particular the response may follow a Poisson distribution or a negative binomial distribution. In addition the model is  able to account for specific phenomena like excess zeros. 

The parameters in model \eqref{eq:basmodel} have an easy interpretation depending on
the function $F(\cdot)$. If one chooses the logistic distribution function one
obtains the conditional transition to higher categories in the form
\[
\log \left(\frac{P(Y_i>r|Y_i\ge r,\xb_i)}{1-P(Y_i>r|Y_i\ge r,\xb_i)}\right) =
\theta_r+\xb_i^T\betab\,,
\]
and the regression coefficients $\beta_1,\hdots,\beta_p$ have the usual interpretation as  in the common binary
logit model. An alternative form is the representation as
continuation ratios
\begin{equation}\label{eq:cr}
\log\left(\frac{P(Y_i>r|\xb_i)}{P(Y_i=r|\xb_i)}\right) = \theta_r+\xb_i^T\betab\,.
\end{equation}
The ratio $P(Y_i>r|\xb_i)/P(Y_i=r|\xb_i)$ compares the probability that the number of counts is larger than $r$ to the probability that the number of counts is equal to $r$. 

The basic assumption of model \eqref{eq:cr} is that the effect of covariates is the same for any given category $r$. This property can also be seen as a form of strict stochastic ordering. That means, if one considers two population that are characterized by the covariate values $\xb$ and $\tilde\xb$, one obtains
\begin{equation}\label{eq:logmodel}
\frac{{P(Y>r|\xb)}/{P(Y=r|\xb)} }{{P(Y>r|\tilde\xb)}/{P(Y=r|\tilde\xb)}} = \exp((\xb-\tilde\xb)^T\betab)\,.
\end{equation}
Thus it is assumed that continuation ratios are the same for each
category $r$.

The modeling of transitions as specified in model \eqref{eq:basmodel} was used in various contexts before.  In ordinal regression transition modeling is known under the name sequential
model; in the logistic version it is called  continuation ratio
model \citep{Agresti:02,TutzBook2011}. Its properties as an ordinal
regression model have been investigated in particular by
\citet{McCullagh:80}. It is also used in discrete survival analysis,
where one parameterizes the discrete hazard function
$\lambda(r|\xb_i)=P(Y_i=r|Y_i\ge r,\xb_i)$ instead of the conditional
probability of transition~\citep{TuSchm2016}.

Modeling of transitions, however, seems not to have been used in the
modeling of count data. The main difference to the use in ordinal
regression and discrete hazard modeling is that, in contrast to
these models, the number of categories is not restricted. In ordinal
models one typically uses up to ten categories, which limits the
number of parameters. In count data, however, the number of
intercepts is unlimited. In extended models, where also the regression coefficients vary across categories (to be considered in Section \ref{sec:tran})
the main problem is the number of parameters that cannot be handled
by simple maximum likelihood estimation. For count data and fixed
predictor value, model (\ref{eq:basmodel}) is a Markov chain model of
order one, because the probability of transition depends only on the
previously obtained category. It is related to
categorical time series, which have been investigated by
\citet{Kaufmann:87}, \citet{FahKau:87}, and \citet{FokKed:2002}.

\subsection{Illustration of Flexibility of the Model}

Before giving details of the fitting procedure we demonstrate the flexibility of the proposed transition model by a small benchmark experiment that was based on 100 replications. We generated samples of size $n=100$ with the outcome values drawn from (i)~a Poisson distribution, $y_i \sim Po(\mu_i=5),\,i=1,\hdots,n$, and (ii)~a negative binomial distribution, $y_i \sim NB(\nu=5/8, \mu_i=5),\,i=1,\hdots,n$, which equals variance $\var(y_i)=45$. 

\begin{figure}[!t]
\centering
\includegraphics[width=0.49\textwidth]{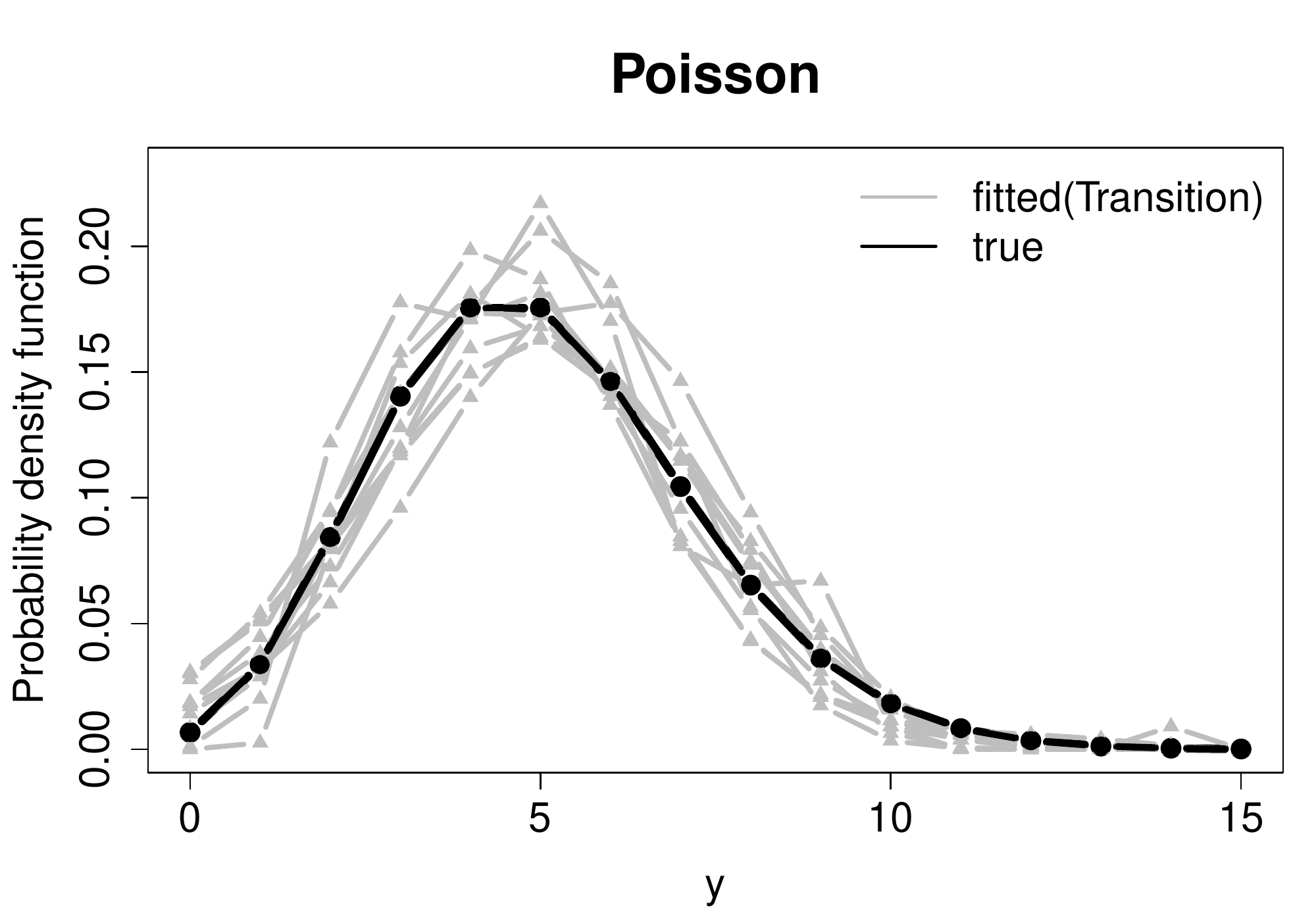}
\includegraphics[width=0.49\textwidth]{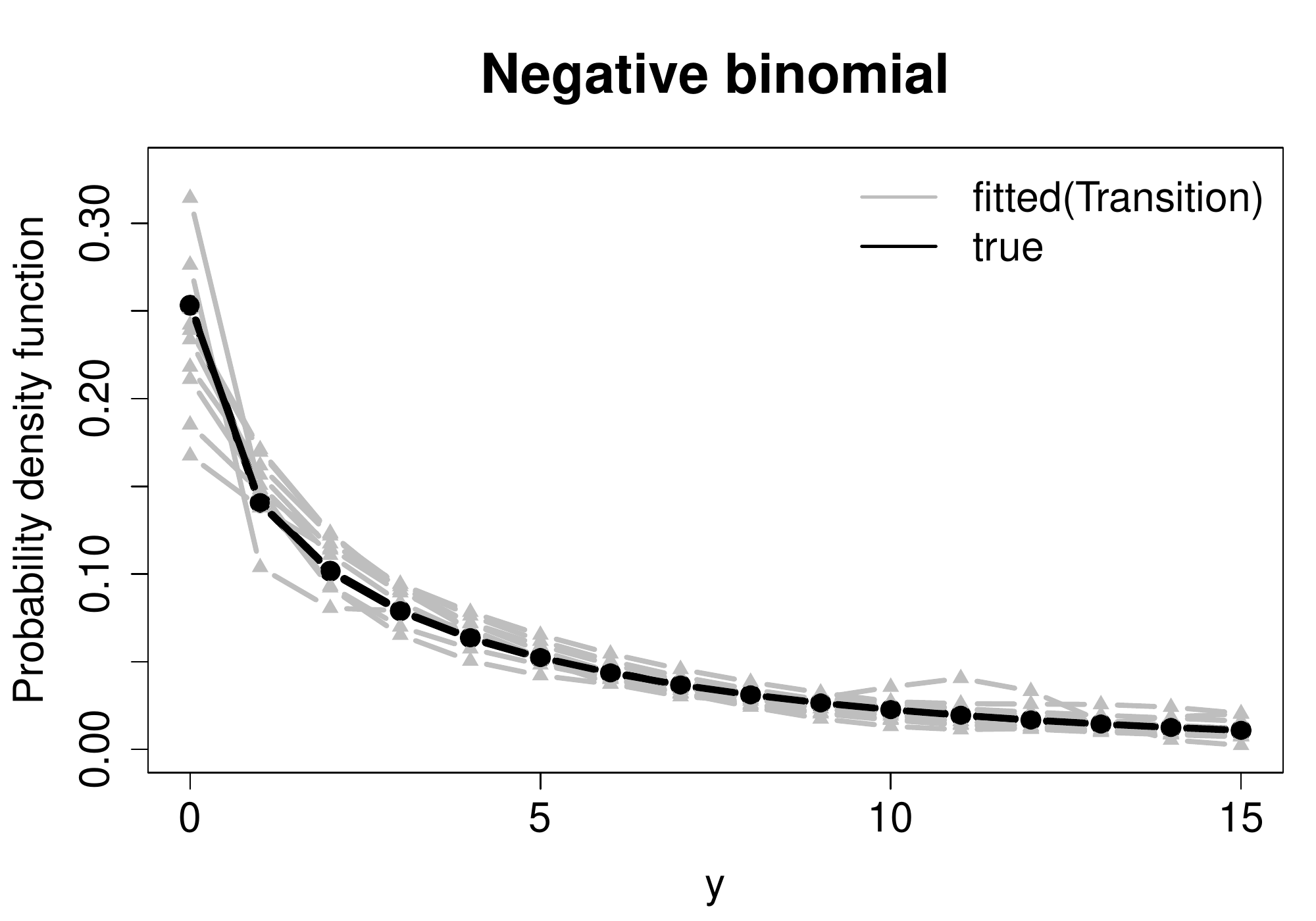}
\includegraphics[width=0.49\textwidth]{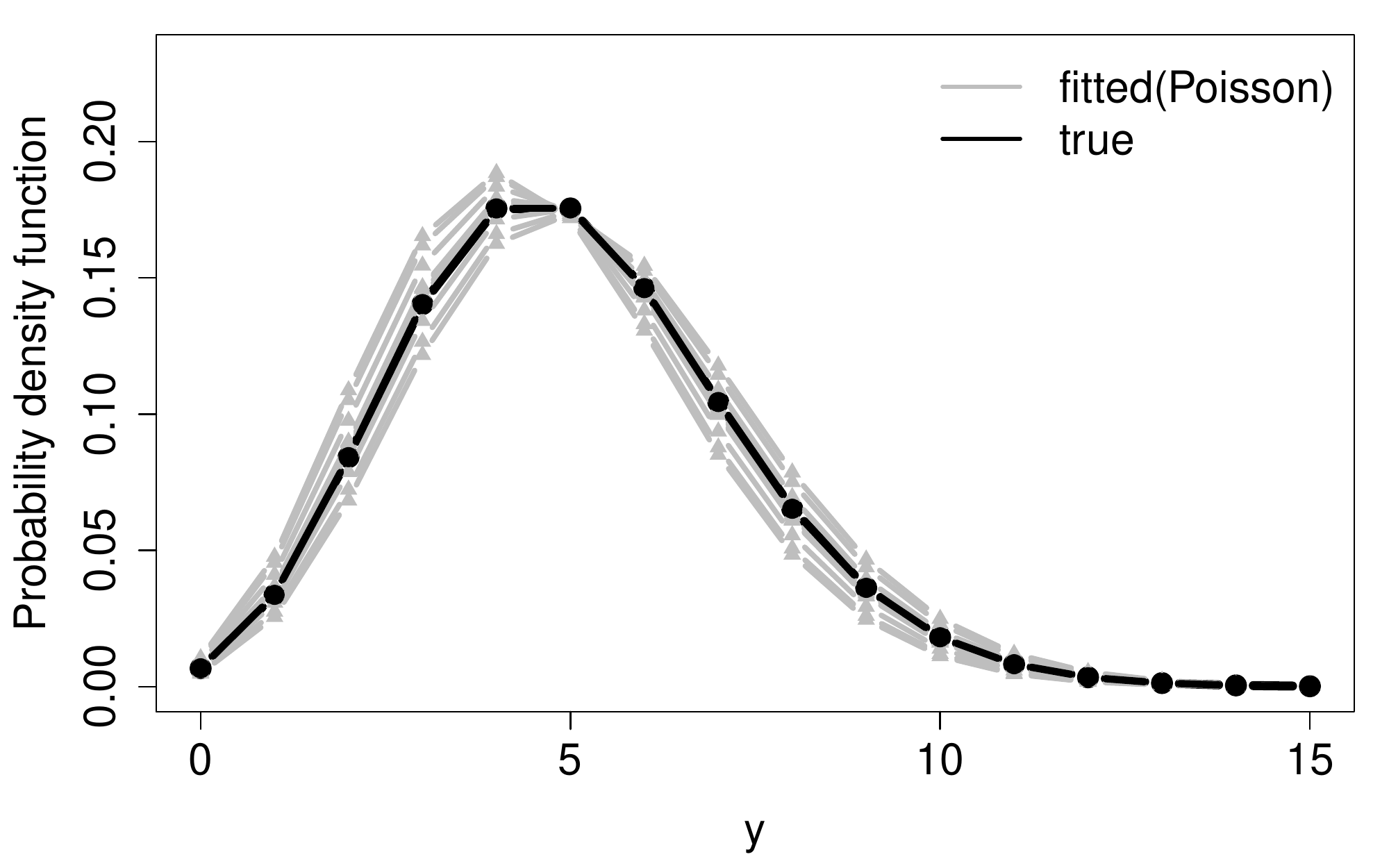}
\includegraphics[width=0.49\textwidth]{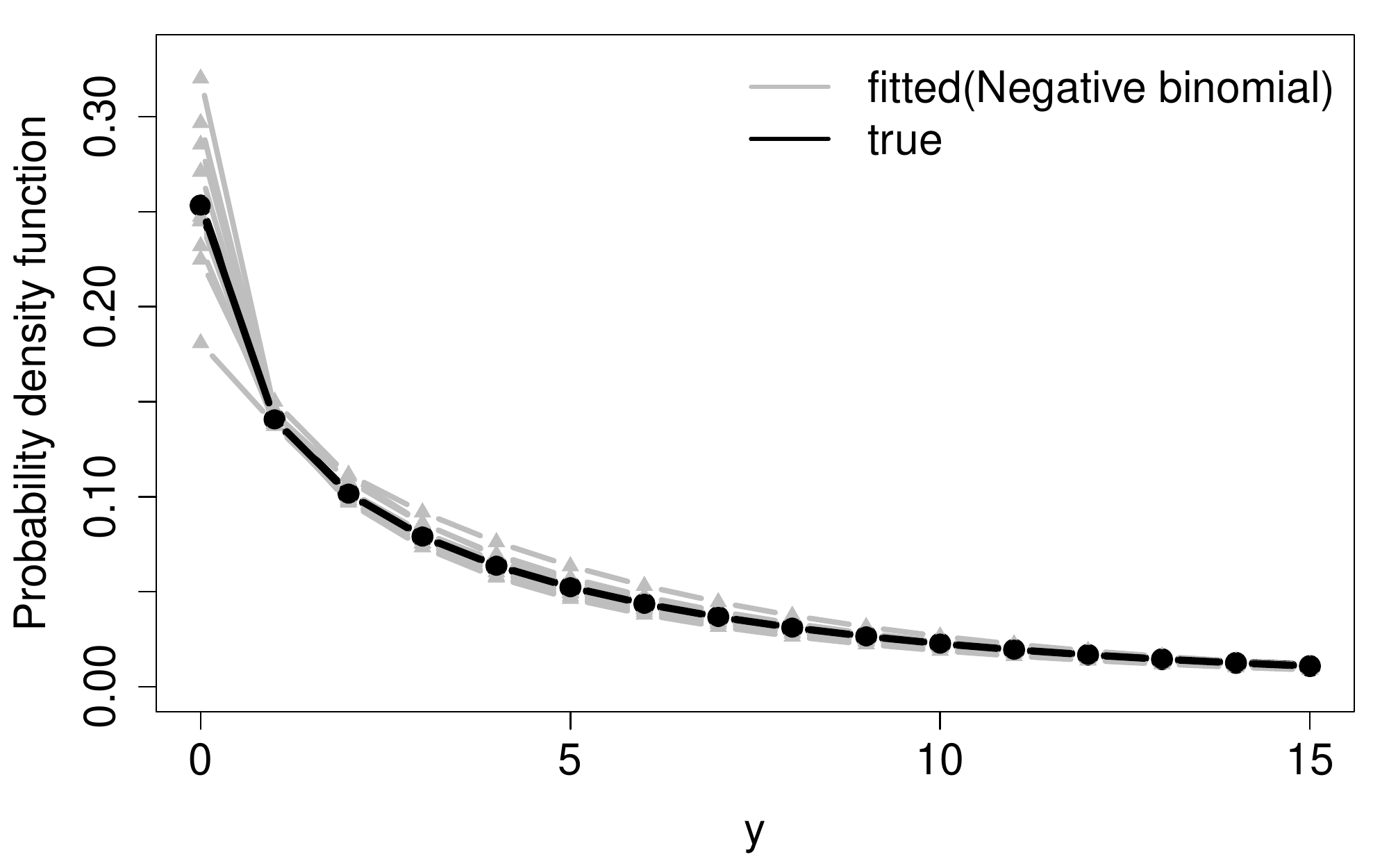}
\includegraphics[width=0.49\textwidth]{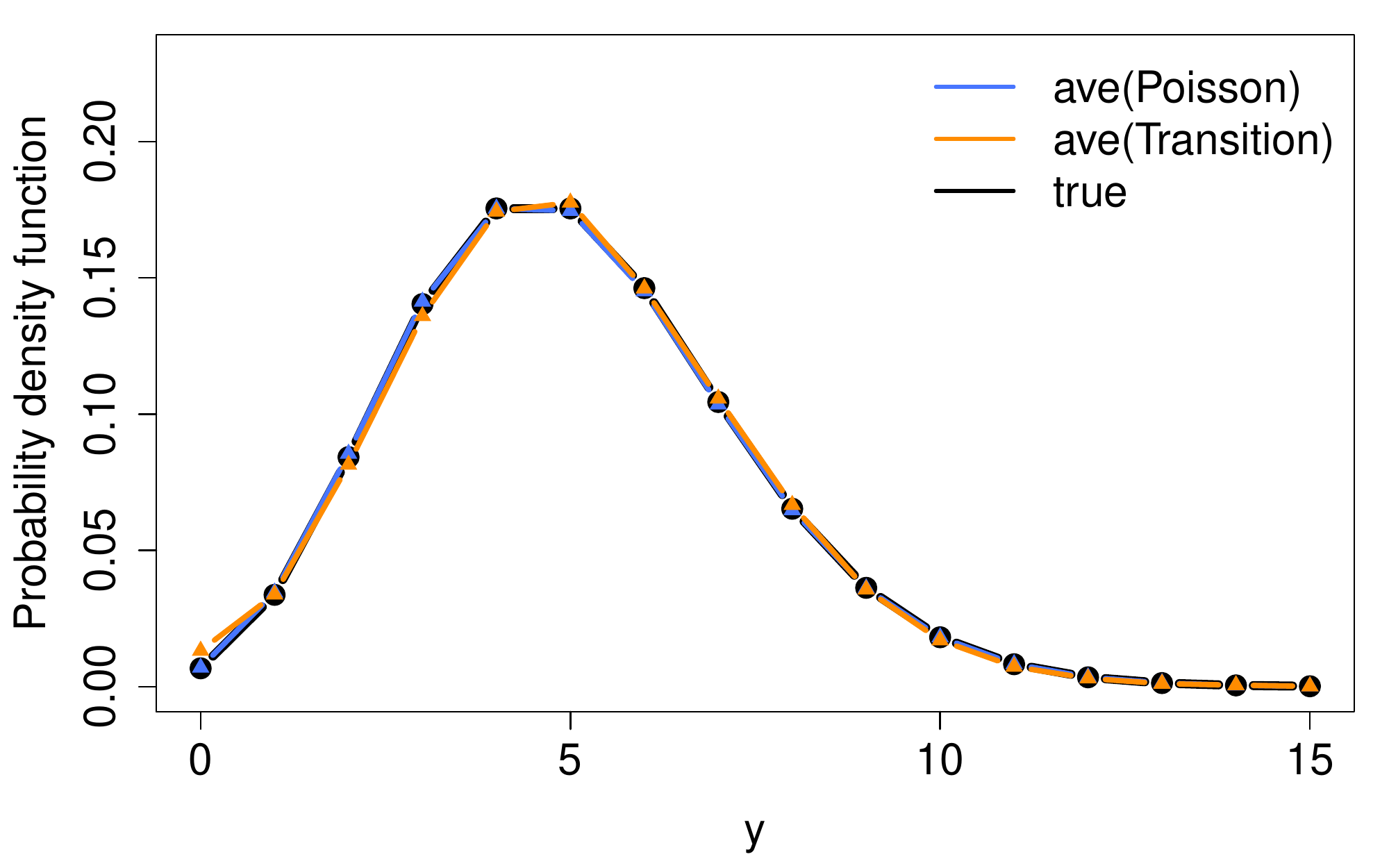}
\includegraphics[width=0.49\textwidth]{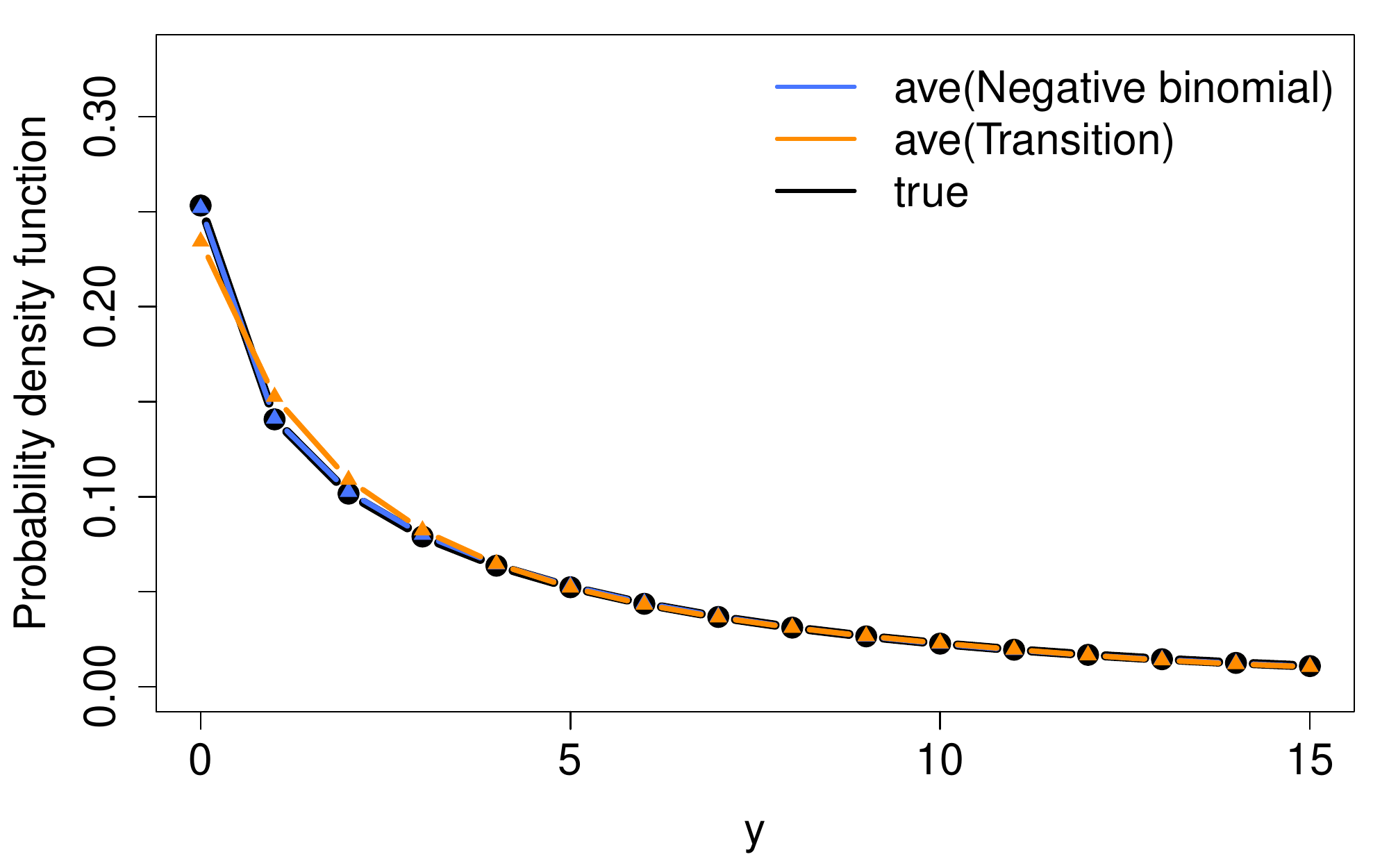}
\caption{Estimated probability density functions when fitting the transition model and the true data-generating model to samples drawn from a Poisson distribution (left) and samples drawn from a negative binomial distribution (right). The upper and middle panels show the estimates of 10 randomly chosen replications when the transition model (upper panel) and the true data generating model (middle panel) are fitted. The lower panel shows the respective average estimates obtained from 100 simulation runs. The black lines correspond to the true probability density function.}
\label{fig:illustration}
\end{figure}

Figure \ref{fig:illustration} shows the estimated probability density functions for 10 randomly chosen replications (upper and middle  panel) and the average estimated probability density function over all 100 replications (lower panel) obtained from fitting the transition model and the true data-generating model (Poisson or negative binomial) to the data, respectively. 
In both cases, it is seen that the transition model is well able to capture the underlying distribution. In particular, the average estimated p.d.f.\,and the true p.d.f.\,(black line) closely coincide for both distributions (lower panel).

\subsection{Estimation}

\subsubsection*{Maximum Likelihood Estimation}

For i.i.d. observations $(Y_i,\xb_i), i=1,\dots,n$, the
log-likelihood has the simple form
 \[
 \ell(\alphab)= \sum_{i=1}^{n} \log (\pi_{ir}),
 \]
where $\alphab^T=(\theta_0, \theta_1, \dots,\betab^T)$ collects all
the parameters and $\pi_{ir}=P(Y_i=r|\xb_i)$ is the probability of
observing category $r$, which for the transition model
(\ref{eq:basmodel}) is given by
\begin{equation}\label{eq:lik}
\begin{aligned}
 \pi_{ir}&=P(Y_i=r|Y_i\ge r, \xb_i)\prod_{s=0}^{r-1}P(Y_i>s|Y_i\ge s, \xb_i)\\
 &=\left(1-F(\theta_r+\xb_i^T\betab)\right)\prod_{s=0}^{r-1}F(\theta_s+\xb_i^T\betab)\,.
\end{aligned}
\end{equation}
For the model considered here it is helpful to represent the
data in a different way. One considers the underlying Markov chain
$Y_{i0}, Y_{i1}, Y_{i2}, \hdots $, where $Y_{ir}= I(Y_i=r)$ with
$I(\cdot)$ denoting the indicator function ($I(a)=1$, if $a$ holds,
$I(a)=0$ otherwise). Then the likelihood can be given in the form
 \[
 \ell(\alphab)= \sum_{i=1}^{n} \sum_{s=0}^{Y_i}Y_{is}\log (\pi_{is})\,.
 \]
By using (\ref{eq:lik}) it can be rewritten as
 \begin{align}\label{eq:likbin}
 \ell(\alphab)= \sum_{i=1}^{n} \sum_{s=0}^{Y_i}Y_{is}\log \left(1-F(\theta_s+\xb_i^T\betab)\right)+(1-Y_{is})F(\theta_s+\xb_i^T\betab)\,.
 \end{align}
In (\ref{eq:likbin}) the realizations of the Markov chain  $Y_{i0},
Y_{i1},\dots, Y_{i,Y_i}$ up to the observed response have the form
$(Y_{i0}, Y_{i1},\dots, Y_{i,Y_i})^T=(0,0,\dots,0,1)$. They can be
seen as dummy variables for the response or as  binary variables
that code if transition to the next category occurred or not. The value $Y_{ir}=0$, $r< Y_i$, denotes that the transition to a higher category than $r$ occurred. If one wants to code transition as 0-1 variable with 1 denoting transition to a higher category one uses
$\tilde{Y}_{ir}=1-Y_{ir}$ yielding the likelihood
 \begin{align}\label{eq:likbin2}
 \ell(\alphab)= \sum_{i=1}^{n} \sum_{s=0}^{Y_i}\tilde Y_{is}F(\theta_s+\xb_i^T\betab)+(1-\tilde Y_{is})\log \left(1-F(\theta_s+\xb_i^T\betab)\right),
 \end{align}
which is obviously equivalent to the log-likelihood of the binary
response model $P(\tilde Y_{ir}=1|\xb_i)= F(\theta_r+\xb_i^T\betab)$ for
observations $\tilde Y_{i0}, \tilde Y_{i1},\dots, \tilde Y_{i,Y_i}$.
Thus the model can be fitted by using maximum likelihood methods for
binary data that encode the sequence of transitions up to
the observed response.

\subsubsection*{Penalized Maximum Likelihood Estimation}

Maximum likelihood (ML) estimators tend to fail because model \eqref{eq:basmodel}  contains  many parameters, in particular the number of intercepts becomes large unless the counts are restricted to very small numbers. Therefore alternative estimators are needed.
We will use penalized maximum likelihood estimates. Then
instead of the log-likelihood \eqref{eq:likbin2} one maximizes the penalized
log-likelihood
\[
\ell_p(\alphab) = \ell(\alphab) - \lambda J(\alphab)\,,
\]
where $\ell(\cdot)$ is the common log-likelihood of the model and
$J(\alphab)$ is a penalty term that penalizes specific structures in
the parameter vector. The parameter $\lambda$ is a tuning parameter
that specifies how serious the penalty term has to be taken. Since
the  intercept parameters $\theta_r$ determine the dimensionality of the estimation problem the penalty is used to
regularize these parameters. 


A reasonable assumption on the $\theta$-parameters is that they are changing slowly over categories. A penalty that enforces smoothing over response categories uses the squared differences between adjacent categories, Let $M_{\text{max}}$ denote the maximal value that has been observed, that is,  $M_{\text{max}}=\max\{Y_i\}$. Then one uses the penalty
\begin{align}\label{eq:quadrdif}
 J(\alphab)= \sum_{s=0}^M(\theta_s-\theta_{s-1})^2\,,
\end{align} 
where $M$ is larger than $M_{\text{max}}$, for example, $M$ can be chosen as the integer closest to $1.2\,M_{\text{max}}$.
When maximizing the penalized log-likelihood one obtains $\theta_{M_{\text{max}}}=\theta_{M_{\text{max}}+1}=\dots=\theta_M$.
It is important to choose $M$ larger than $M_{\text{max}}$ to account for possibly larger future observations and to avoid irregularities at the boundaries. If one uses $\lambda=0$ in (\ref{eq:quadrdif}) one obtains the ML estimates. In the extreme case $\lambda \rightarrow\infty$ all parameters obtain the same value.

An alternative, more general smoothing technique uses penalized splines as
proposed by \citet{EilMar:96}. Let the $\theta$-parameters be
specified as a smooth function over categories by using an expansion
in basis functions of the form
\[
\theta_r=\sum_{k=1}^m\gamma_k\phi_k(r)\,,
\]
where $\phi_k(\cdot)$ are fixed basis functions. 
We will use B-splines \citep{EilMar:96} 
on equally spaced knots in the range $[0,M]$, where $M$ is again a larger value than the maximal observed response. The
penalty now does not refer to the $\theta$-parameters themselves but
to the $\gamma$-parameters.  A flexible form is
\begin{equation}\label{eq:DIFF}
J(\alphab)=\sum_{k=d+1}^m(\Delta^d \gamma_k)^2\,,
\end{equation}
where $\Delta^d$ is the difference operator, operating on adjacent
B-spline coefficients, that is,
$\Delta\gamma_k=\gamma_k-\gamma_{k-1},
\Delta^2\gamma_k=\Delta(\gamma_k-\gamma_{k-1})=\gamma_k-2\gamma_{k-1}+\gamma_{k-2}$.
The method is referred to as \textit{P-splines} (for penalized
splines). 


\subsubsection*{Embedding into the Framework of Varying-Coefficients Models }
The transition variables 
$(\tilde Y_{i0}, \tilde Y_{i1},\dots, \tilde Y_{i,Y_i})^T=(1,1,\dots,1,0)$ in \eqref{eq:likbin2} are binary variables. The log-likelihood is the same as for binary response models of the form
$P(\tilde Y_{ir}=1|\xb_i)=F(\theta_r+\xb_i^T\betab)$. The binary models can also be seen as varying-coefficients models of the form  
$P(\tilde Y_{ir}=1|\xb_i)=F\left(\beta(r)+\xb_i^T\betab\right)$, where $\beta(r)=\theta_r$ is an unknown function of the category, that is, the intercepts vary across categories. By considering the category as an explanatory variable one may treat the models as specific varying-coefficients models and use existing software for fitting (see Section \ref{sec:software} for further details). 

\subsection{Selection of Smoothing Parameter and Prediction Accuracy}

For the choice of the tuning parameter (for example by resampling or cross-validation) a
criterion for the accuracy of prediction is needed. A classical
approach in linear models is to estimate the mean and compare it to
the actually observed response by using the quadratic distance. But
since the whole distribution given a fixed covariate is estimated it
is more appropriate to compare the estimated distribution to the
degenerated distribution that represents the actual observation by
using loss functions.

Candidates for loss functions are the quadratic loss function
$L_2(\pib_i,\hat{\pib}_i)=\sum\nolimits_r(\pi_{ir}-\hat{\pi}_{ir})^2$ and
the Kullback-Leibler loss $L_{KL}(\pib_i,\hat{\pib}_i)=\sum\nolimits_r\pi_{ir}\log(\pi_{ir}/\hat{\pi}_{ir})$, where the vectors $\pib_i=(\pi_{i0},\pi_{i1},\hdots)^\top$ and $\hat\pib_i=(\hat\pi_{i0},\hat\pi_{i1},\hdots)^\top$ denote the true and estimated probabilities. When the true probability vector is
replaced by the 0-1 vector of observations $\Yb_i=(Y_{i0},Y_{i1},Y_{i2},\hdots)^\top$ one obtains for the quadratic loss the Brier score
\[
L_2(\Yb_i,\hat\pib_i)  = (1-\hat\pi_{i,Y_i})^2+\sum_{r\neq Y_i}{}{\hat\pi_{i,r}^2}\,.
\]
For the Kullback-Leibler loss one obtains the logarithmic score
$L_{KL}(\Yb_i,\hat\pib_i) = -\log(\hat\pi_{i,Y_i})$. The latter  has the
disadvantage that the predictive distribution $\hat{\pib}_i$ is only
evaluated at the observation $Y_i$. Therefore, it does not take the
whole predictive distribution into account. As
\citet{Gneitingetal:2007} postulate, a desirable predictive
distribution should be as sharp as possible and well calibrated.
Sharpness refers to the concentration of the distribution and
calibration to the agreement between the distribution and the
observation. For  count data, a more appropriate loss function
derived from the continuous ranked probability score
\citep{Gneitingetal:2007}, which will also be used in the applications in the next section, is
\begin{equation}\label{eq:RPSc}
L_{RPS}(Y_i,\hat{\pib}_i)=\sum_r\left(\hat\pi_i(r)- I(Y_i \le r)\right)^2,
\end{equation}
where $\hat\pi_i(r)=\hat\pi_{i0}+\hat\pi_{i1}+\hdots+\hat\pi_{ir}$ represents the
cumulative distribution. It was also used by
\citet{czado2009predictive} for the predictive assessment of count
data.


\section{Applications} \label{sec:app}

To illustrate the applicability and the added value of the proposed transition model, we show the results of three real-world data examples comparing the various approaches introduced in the previous sections. Specifically, we consider 
\begin{itemize}
\setlength{\itemsep}{0cm}
\item[(i)] the Poisson model, hereinafter referred to as \textit{Poisson}, 
\item[(ii)] the negative binomial model, referred to as \textit{NegBin},
\item[(iii)] the zero-inflation model \eqref{eq:Zero} using a Poisson model for the responders and a logit model to determine the class membership, referred to as \textit{Zero}, 
\item[(iv)] the hurdle model \eqref{eq:hurdle} using a logit model for $f_1$ and a Poisson model for $f_2$, referred to as \textit{Hurdle},
\item[(v)] the transition model with a quadratic difference penalty \eqref{eq:quadrdif} on the intercepts, referred to as \textit{QuadPen} and  
\item[(vi)] the transition model, where the intercepts are expanded in cubic B-splines using a first order difference penalty \eqref{eq:DIFF}, referred to as \textit{P-Splines}. 
\end{itemize}
To determine the optimal smoothing parameters $\lambda$ for models (v) and (vi) and to compare the predictive performance of all six approaches we used the ranked probability score \eqref{eq:RPSc}. For this purpose, we repeatedly (100 replications) generated subsamples without replacement containing $2/3$ of the observations in the original data and computed the ranked probability score from the remaining test data sets (i.e., from 1/3 of the original data). 

\subsection{Absenteeism from School}

We first consider a sociological study on children in Australia. The data set is available in the R package \textbf{MASS} \citep{MASS} and was initially analysed by \citet{aitkin1979}. The data consists of a sample of 146 children from New South Wales, Australia. The outcome of interest is the number of days a child was absent from school in one particular school year, taking values $Y_i \in \{0,\hdots,81\}$. The covariates included in the models are Aboriginal ethnicity (Eth; 0:~no, 1:~yes), gender (Sex; 0:~female, 1:~male), the educational stage (Edu; 0:~primary, 1:~first form, 2:~second form, 3:~third form) and a learner status (Lrn; 0:~average, 1:~slow). 

\begin{figure}[!t]
\centering
\includegraphics[width=0.75\textwidth]{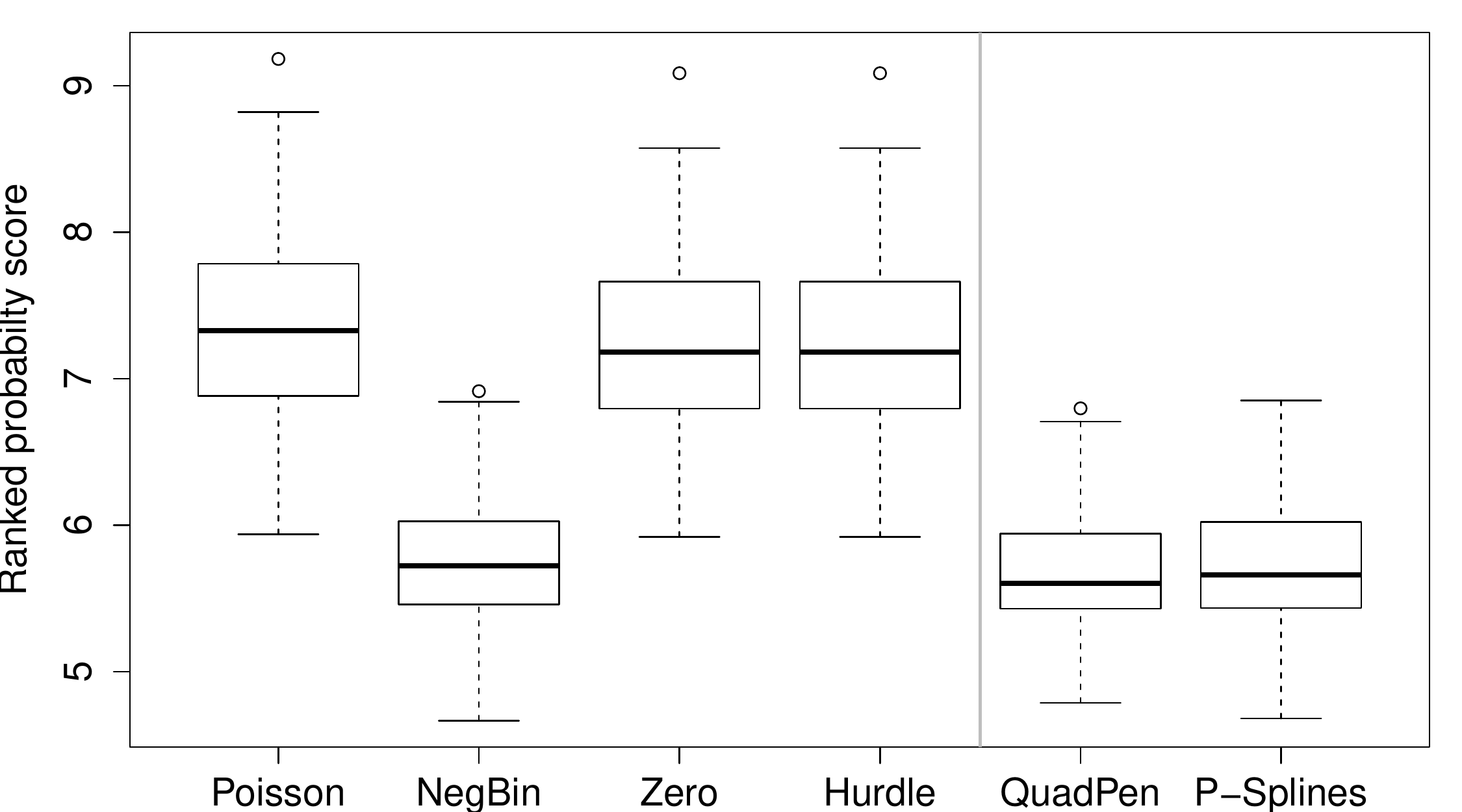}
\caption{Analysis of the absenteeism from school data. The boxplots show the ranked probability score (for $r \in \{0,\hdots,30\}$) of the four classical models (left) and the two transition models (right). All methods were fitted to 100 subsamples without replacement of size $100$ and evaluated on the remaining $46$ observations each.}
\label{fig:rps_quine}
\end{figure}

Figure \ref{fig:rps_quine} shows the results of the resampling experiment comparing the six different approaches. The ranked probability score was computed from the test data sets (containing 46 observations each) for outcome values $r \in \{0,\hdots,30\}$. This range of values was chosen to ensure outcome values up to $r=30$ in each of the learning and test samples. It is seen that the negative binomial model (second boxplot) and the two transition models (fifth and sixth boxplot) outperform  the three alternative approaches with only minor differences between them. This result indicates that the negative binomial distribution is much more appropriate than the Poisson distribution for the analysis of this data set. Importantly, the performance of the flexible transition model in terms of accuracy of prediction is the same as for the negative binomial model. The transition model is well able to capture the essential characteristics of the data. 

\begin{table}[!t]
\centering
\begin{tabularsmall}{lrrrcrrr}
  \toprule
	&\multicolumn{3}{c}{\bf{NegBin}}&&\multicolumn{3}{c}{\bf{P-Splines}}\\
 & coef & se & z-value && coef & se & z-value \\ 
  \midrule
  \bf{Eth} & 0.569 & 0.153 & 3.713 && 0.585 & 0.178 & 3.291 \\ 
  \bf{Sex} & 0.082 & 0.160 & 0.515 && 0.082 & 0.185 & 0.446 \\  
  \bf{Edu:1} & -0.448 & 0.240 & -1.870 && -0.470 & 0.266 & -1.764 \\
  \bf{Edu:2} & 0.088 & 0.236 & 0.373 && 0.087 & 0.271 & 0.321 \\  
  \bf{Edu:3} & 0.357 & 0.248 & 1.437 && 0.368 & 0.277 & 1.329 \\
  \bf{Lrn} & 0.292 & 0.186 & 1.566 && 0.309 & 0.205 & 1.507 \\ 
   \bottomrule
\end{tabularsmall}
\caption{Analysis of the absenteeism from school data. Parameter estimates, standard errors and z-values obtained from fitting the negative binomial model~(left) and the transition model with penalized B-splines~(right).} 
\label{tab:coef_quine}
\end{table}


The estimated coefficients $\hat{\betab}$ and the corresponding standard errors and z-values obtained by the negative binomial model and the transition model with penalized B-splines using the whole sample of $n=146$ children are given in Table~\ref{tab:coef_quine}. Overall, the results of both models widely coincide. From the z-values it can be derived that only ethnicity has a significant effect on the outcome (at the 5\% type 1 error level). Based on the negative binomial model, the expected number of days a child is absent from school is increased by the factor $\exp(0.569)=1.766$ for the group of aboriginal children. In terms of the transition model, the continuation ratio (defined in \eqref{eq:logmodel}) is increased by the factor $\exp(0.585)=1.795$, indicating higher counts in the group of aboriginal children. 

\subsection{Demand for Medical Care } 
\begin{figure}[!t]
\centering
\includegraphics[width=0.8\textwidth]{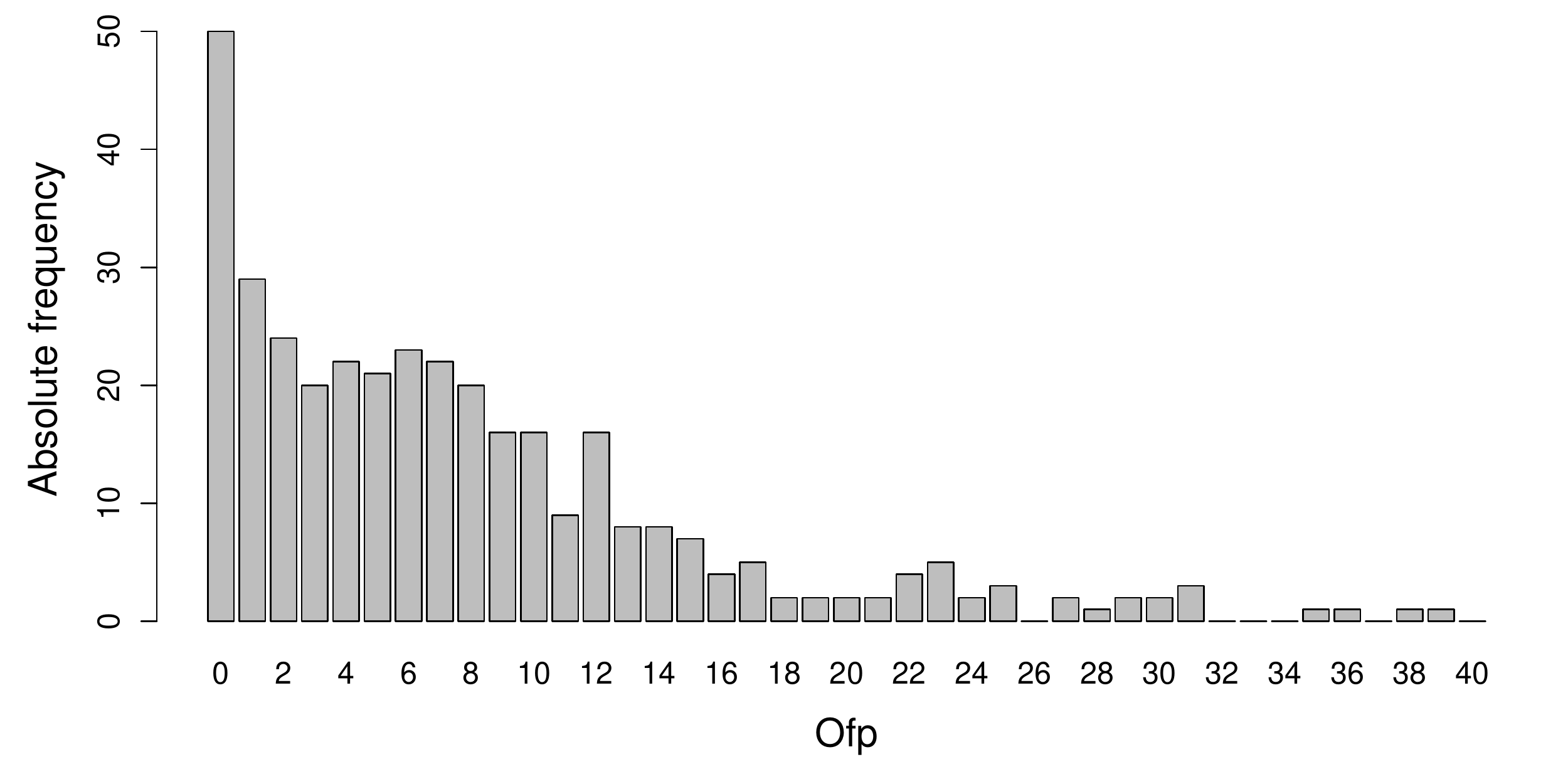}
\caption{Distribution of the outcome variable Ofp measured in the National Medical Expenditure survey ($n=356$).}
\label{fig:counts_ofp}
\end{figure}

\citet{DebTri:97} analyzed the demand for medical care for individuals, aged 66 and over, based
on a dataset from the U.S. National Medical Expenditure survey in
1987/88. The data (``NMES1988'') are available from the R package~\textbf{AER} \citep{KleZei:2008}. Like \citet{Zeietal:2008} we consider the
number of physician/non-physician office and hospital outpatient
visits (Ofp) as outcome variable. The covariates used in the
present analysis are the self-perceived health status (Health; 0: poor, 1: excellent), the number of hospital stays (Hosp), the number of
chronic conditions (Numchron), age, maritial status (Married; 0: no, 1: yes), and number of
years of education (School). Since the effects vary across gender, we restrict consideration  to male patients ($n=356$). Figure \ref{fig:counts_ofp} shows the unconditional distribution of the outcome $Y_i \in \{0,\hdots,40\}$. 

\begin{figure}[!t]
\centering
\includegraphics[width=0.75\textwidth]{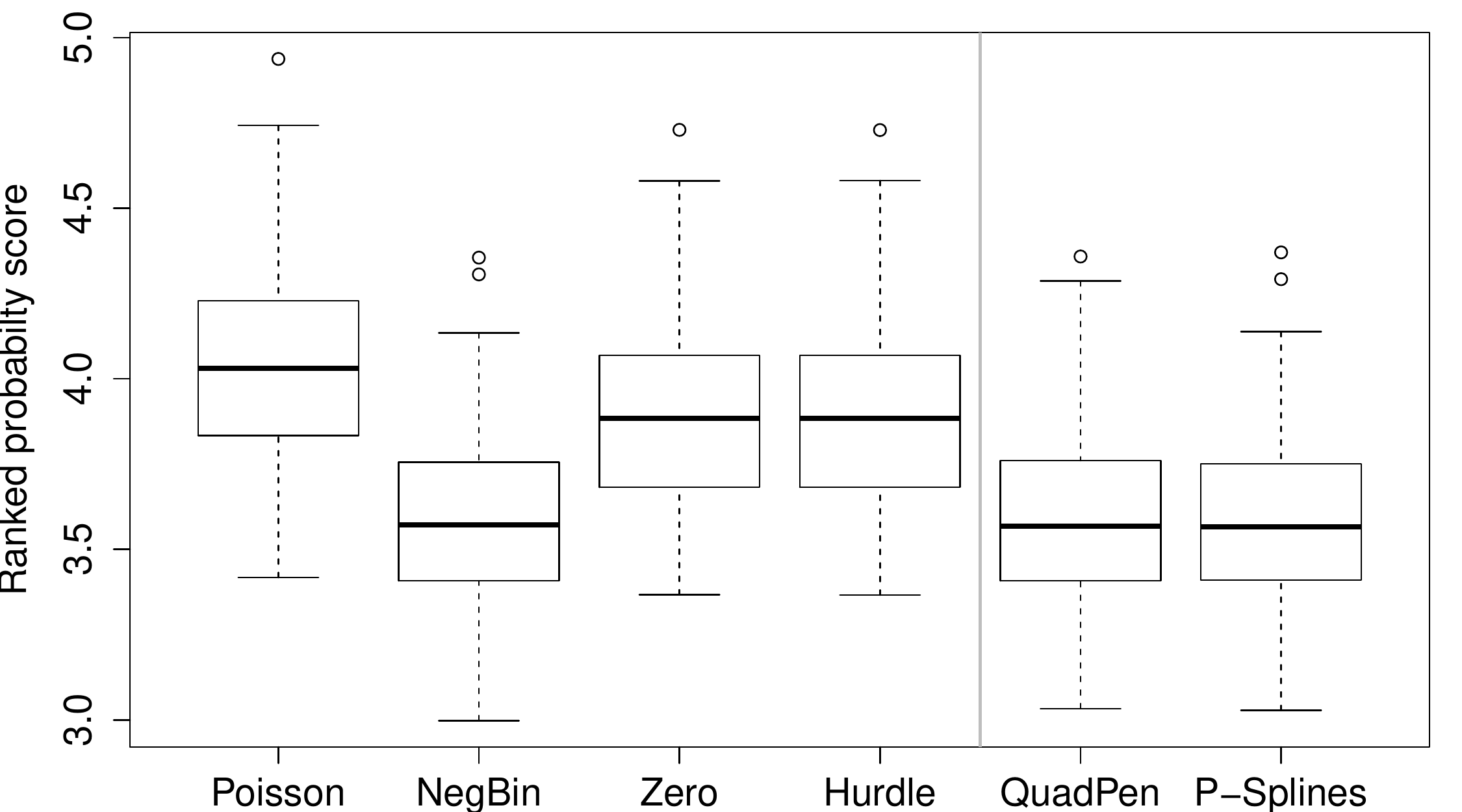}
\caption{Analysis of the medical care data. The boxplots show the ranked probability score (for $r \in \{0,\hdots,30\}$) of the four classical models (left) and the two transition models (right). All methods were fitted to 100 subsamples without replacement of size $237$ and evaluated on the remaining $119$ observations each.}
\label{fig:rps_ofp}
\end{figure}

The ranked probabilty scores for outcome values $r\in \{0,\hdots,30\}$ obtained from the approaches (i) to (vi) are shown in Figure \ref{fig:rps_ofp}. When fitting the transition models the minimal ranked probability score (averaged over 100 replications) was obtained for $\lambda=5$ (QuadPen) and $\lambda=16$ (P-Splines), respectively. Similar to the previous example the negative binomial model (second boxplot) and the two transition models (fifth and sixth boxplot) performed considerably better than the Poisson model and the two models accounting for excess zeros.

\begin{table}[!t]
\centering
\begin{tabularsmall}{lrrrcrrr}
  \toprule
	&\multicolumn{3}{c}{\bf{NegBin}}&&\multicolumn{3}{c}{\bf{P-Splines}}\\
 & coef & se & z-value && coef & se & z-value \\ 
  \midrule
  \bf{Health} & -0.681 & 0.140 & -4.863 && -0.794 & 0.158 & -5.021 \\ 
  \bf{Hosp} & 0.164 & 0.053 & 3.092 && 0.197 & 0.068 & 2.878 \\ 
  \bf{Numchron} & 0.058 & 0.040 & 1.452 && 0.057 & 0.043 & 1.312 \\ 
  \bf{Age} & 0.024 & 0.080 & 0.300 && 0.031 & 0.094 & 0.325 \\ 
  \bf{Married} & 0.074 & 0.122 & 0.608 && 0.067 & 0.138 & 0.482 \\ 
  \bf{School} & 0.042 & 0.013 & 3.259 && 0.045 & 0.014 & 3.127 \\ 
   \bottomrule
\end{tabularsmall}
\caption{Analysis of the medical care data. Parameter estimates, standard errors and z-values obtained from fitting the negative binomial model~(left) and the transition model with penalized B-splines~(right).} 
\label{tab:coef_ofp}
\end{table}

\begin{figure}[!t]
\centering
\includegraphics[width=0.45\textwidth]{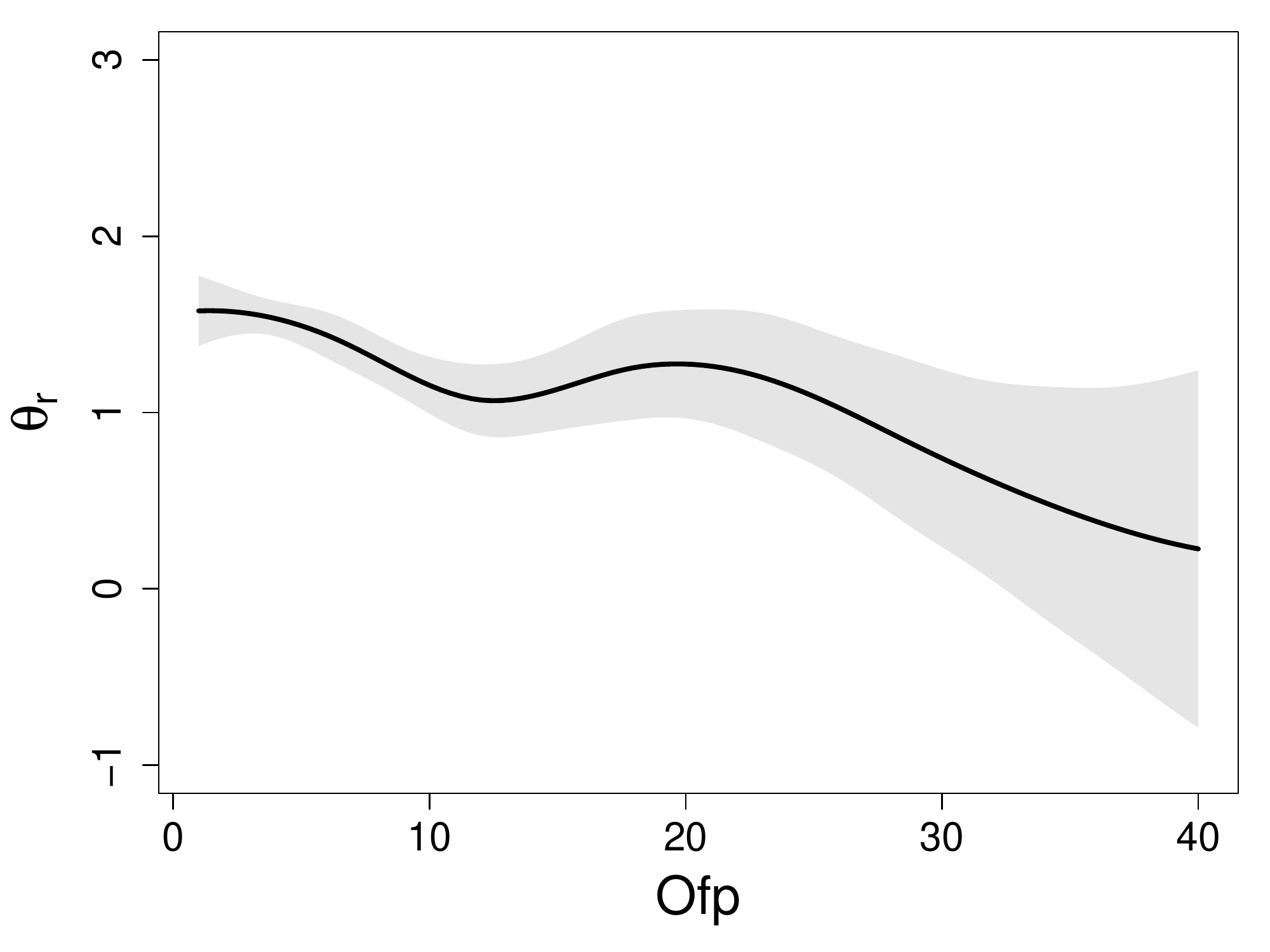}
\caption{Analysis of the medical care data. Smooth function of the $\theta$-parameters obtained from fitting the transition model with penalized B-splines~($\lambda=16$).}
\label{fig:theta_ofp}
\end{figure}

The estimated coefficients $\hat{\betab}$ and the corresponding standard errors and z-values obtained by the negative binomial model and the transition model with penalized B-splines using the whole sample of $n=356$ patients are given in Table~\ref{tab:coef_ofp}. Again, the two models yielded very similar results. An excellent health status reduced the expected number of visits, whereas the number of hospital stays and the number of years of education significantly increased the expected number of visits. 
Figure \ref{fig:theta_ofp} shows the fitted smooth function of the $\theta$-parameters obtained by the transition model with penalized B-splines, which represents the basic distribution of the counts. The function reveals decreasing coefficients with a local peak at $\sim 20$ visits.

\section{Transition Model with Varying Coefficents}\label{sec:tran}

An extended form of the transition model assumes for $Y_i \in \{0,1,2,\hdots\}$
\begin{equation}\label{eq:basmodelext}
P(Y_i>r|Y_i\ge r,\xb_i)= F(\theta_r+\xb_i^T\betab_r)\,,
\end{equation}
where the regression coefficients  $\betab_r^T=(\betab_{1r},\dots,\betab_{pr})$ may vary over
categories. The parameter $\betab_{jr}$ represents the weight on
variable $j$ for the transition to higher categories than $r$. Within the basis functions approach each $\beta$-parameter can be represented as 
\[
\betab_{jr}=\sum_{k=1}^m\gamma_{jk}\phi_k(r)\,,\quad j=1,\hdots,p\,,
\]
where $\phi_k(r)$ are fixed basis functions. Then the whole predictor of the model becomes 
\[
\log\left(\frac{P(Y_i>r|\xb_i)}{P(Y_i=r|\xb_i)}\right) = \sum_{k=1}^m\gamma_k\phi_k(r)+ \sum_{j=1}^p\sum_{k=1}^m {x_{ij}\gamma_{jk}\phi_k(r)}\,.
\]

\subsubsection*{Extended Model for Excess Zeros}
A specific model with varying coefficients that is tailored to the case of excess zeros is obtained if one separates the first transition from all the other transitions. In the model 
\begin{equation}\label{eq:specific}
\begin{aligned} 
&P(Y_i>0|\xb_i )= F(\theta_0+\xb_i^T\betab_0)\,,\\
&P(Y_i>r|Y_i\ge r,\xb_i)= F(\theta_r+\xb_i^T\betab)\,, \quad r=1,2,\hdots\,,
\end{aligned}
\end{equation}
the first transition is determined by the parameter vector $\betab_0$ while the other transitions are determined by the parameter vector $\betab$. As in zero-inflated count models and hurdle models one specifies separate effects that model the occurrence of excess zeros. 

In Model \eqref{eq:specific} one has varying coefficients $\beta(r)=\theta_r$ in the second equation of the model that vary across categories $r=1,2,\hdots$. These can again be fitted using a quadratic difference penalty or penalized B-splines as described for the basic model in Section~\ref{sec:transition}. 

\subsection{Absenteeism from School (contd.)}

\begin{figure}[!t]
\centering
\includegraphics[width=0.95\textwidth]{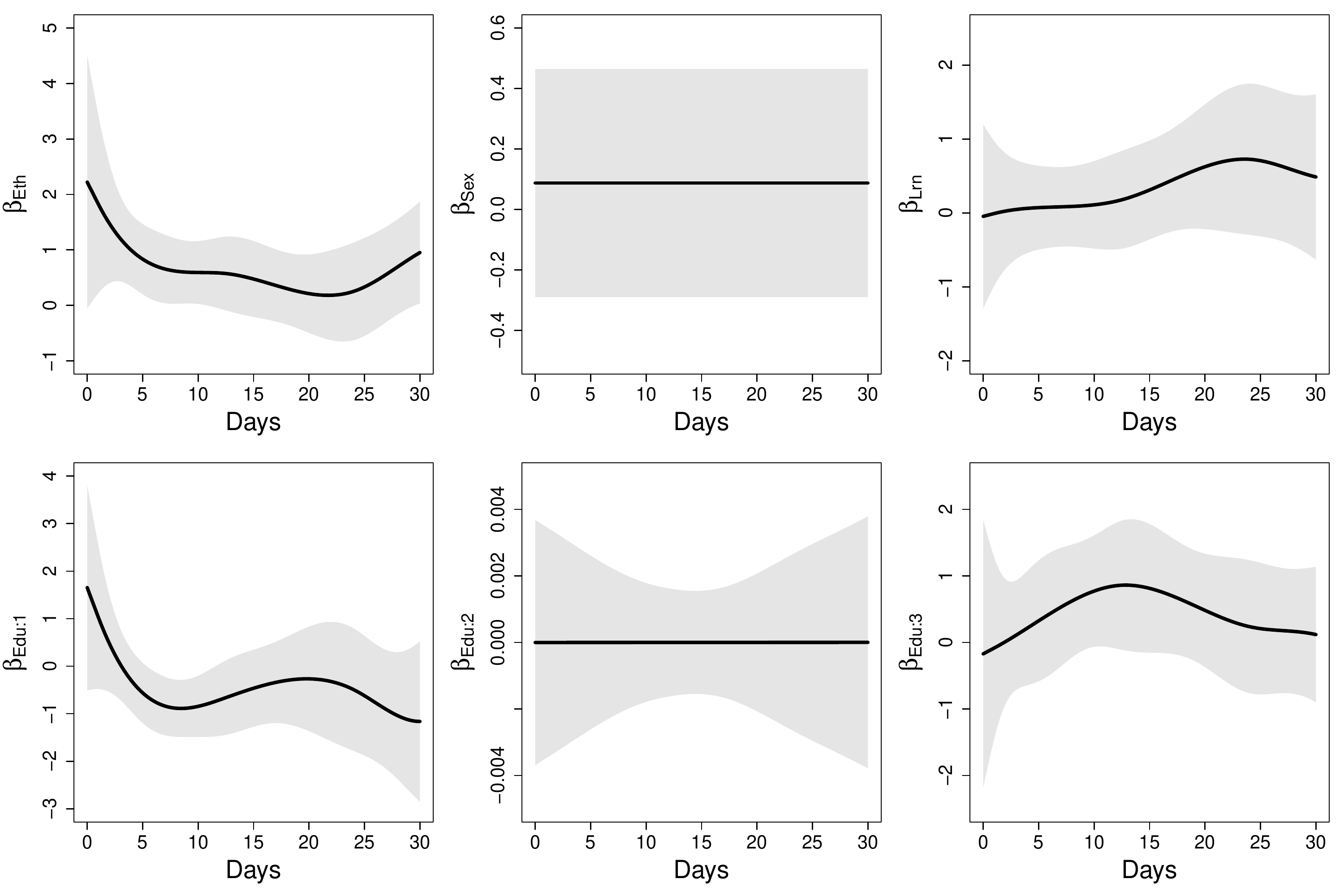}
\caption{Analysis of the absenteeism from school data. Smooth estimates~$\betab_r$~(for $r \in \{0,\hdots,30\}$) obtained from fitting the extended transition model with varying coefficients in all covariates.}
\label{fig:smooth_quine}
\end{figure}

Let us again consider the study on schoolchildren in Australia. Figure \ref{fig:smooth_quine} shows the results when fitting the extended transition model \eqref{eq:basmodelext} in its most general, that means the coefficients of all covariates are expanded in cubic B-splines using a first order difference penalty.

As in the basic transition model (cf. Table \ref{tab:coef_quine}) the estimated functions indicate non-significant constant effects for girls compared to boys and for second form pupils compared to primary pupils (Edu:2). However, there are significant non-linear effects for ethnicity (which was also significant in the basic model) as well as for first form pupils (Edu:1) and learner status.  

As an example, let us consider how  slow learners compare to average learners (upper right panel of Figure \ref{fig:smooth_quine}). It is seen that the continuation ratio increases by the factor $\exp(0.024)=1.024$ ($r=0$) up to the factor $\exp(0.726)=2.067$ ($r=23$), which doubles the probability for a higher count. That means the type of learner has a  stronger impact on the days of absence if the student already has been absent for many days.

\subsection{Demand for Medical Care (contd.)}

\begin{figure}[!t]
\centering
\includegraphics[width=0.45\textwidth]{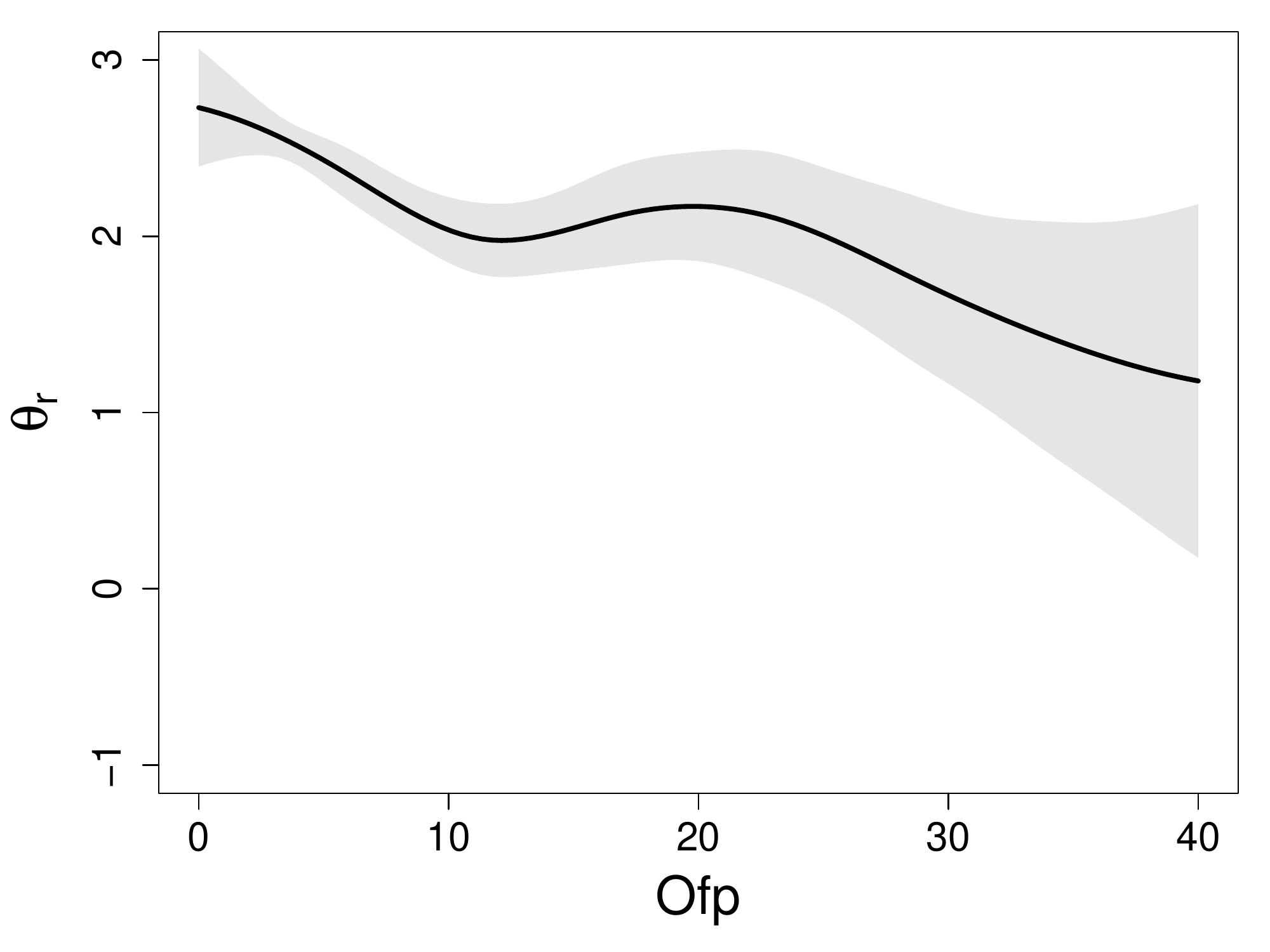}
\caption{Analysis of the medical care data. Smooth function of the $\theta$-parameters obtained from fitting the extended transition model with penalized B-splines~($\lambda=16$).}
\label{fig:theta_ofp_zero}
\end{figure}

The results from fitting the extended model \eqref{eq:specific} accounting for excess zeros to the data of the National Medical Expenditure survey are shown in Table \ref{tab:coef_ofp_zero} and Figure \ref{fig:theta_ofp_zero}. There are remarkable differences compared to the previous results in Table \ref{tab:coef_ofp}: (i) the number of years of education (school) had a significant effect on the first transition, but there was no evidence for an effect on the other transitions ($z$-value = 1.591), (ii) the number of chronic conditions, which was not significant in the basic model, showed a significant positive effect ($\hat{\beta}_0=0.591$) on the first transition (driving the decision to consult a doctor or not), and (iii) an excellent health status and the number of hospital stays had significant effects only in  part of the model that models the transition to higher categories given the number of visits was already above zero. 

The ranked probability score of the extended model (evaluated on the test data sets and averaged over 100 replications) was $3.562$, which indicates the best predictive performance among all considered models (see also Figure \ref{fig:rps_ofp}).

\begin{table}[!t]
\centering
\begin{tabularsmall}{lrrrcrrr}
  \toprule
	&\multicolumn{3}{c}{\bf{Zero}}&&\multicolumn{3}{c}{\bf{Non-Zero}}\\
 & coef & se & z-value && coef & se & z-value \\ 
  \midrule
	$\thetab_0$& -6.421&2.314&-2.774&&---&---&---\\
  \bf{Health} &  -0.375 & 0.466 & -0.806 && -0.749 & 0.171 & -4.379 \\ 
  \bf{Hosp} & 0.608 & 0.373 & 1.631 && 0.178 & 0.069 & 2.574 \\  
  \bf{Numchron} & 0.591 & 0.187 & 3.161 && 0.015 & 0.045 & 0.336 \\  
  \bf{Age} & 0.394 & 0.270 & 1.456 && -0.045 & 0.102 & -0.440 \\  
  \bf{Married} & 0.512 & 0.361 & 1.417 && -0.007 & 0.152 & -0.044 \\ 
  \bf{School} & 0.161 & 0.041 & 3.918 && 0.025 & 0.016 & 1.591 \\ 
   \bottomrule
\end{tabularsmall}
\caption{Analysis of the medical care data. Parameter estimates, standard errors and z-values obtained from fitting the extended transition with penalized B-splines accounting for excess zeros.} 
\label{tab:coef_ofp_zero}
\end{table}

\subsection{Boating Trips} 

As third example we consider data based on a survey in $1980$ administered to $n=659$ leisure boat owners in eastern Texas, which is available from R package~\textbf{AER} \citep{KleZei:2008} and was analyzed before by \citet{ozuna1995}. Here, the outcome of interest is the number of recreational boating trips to Lake Somerville, $Y_i \in \{0,\hdots,40\}$. Note that $417$ (63\%) observations take the value zero, which calls for a model accounting for excess zeros (two outliers with outcome values 50 and 88 were excluded). 

The five covariates used in the present analysis are the facility's subjective quality ranking (Quality; 1: very negative -- 5: very positive), an indicator, if the individual did water-skiing at the lake (Ski; 0: no, 1:yes), the annual household income (Income; in 1,000 USD), an indicator, if the individual payed an annual user fee at the lake (Userfee; 0: no, 1: yes) and the expenditure when visiting the lake (Cost; in USD). 

\begin{figure}[!t]
\centering
\includegraphics[width=0.75\textwidth]{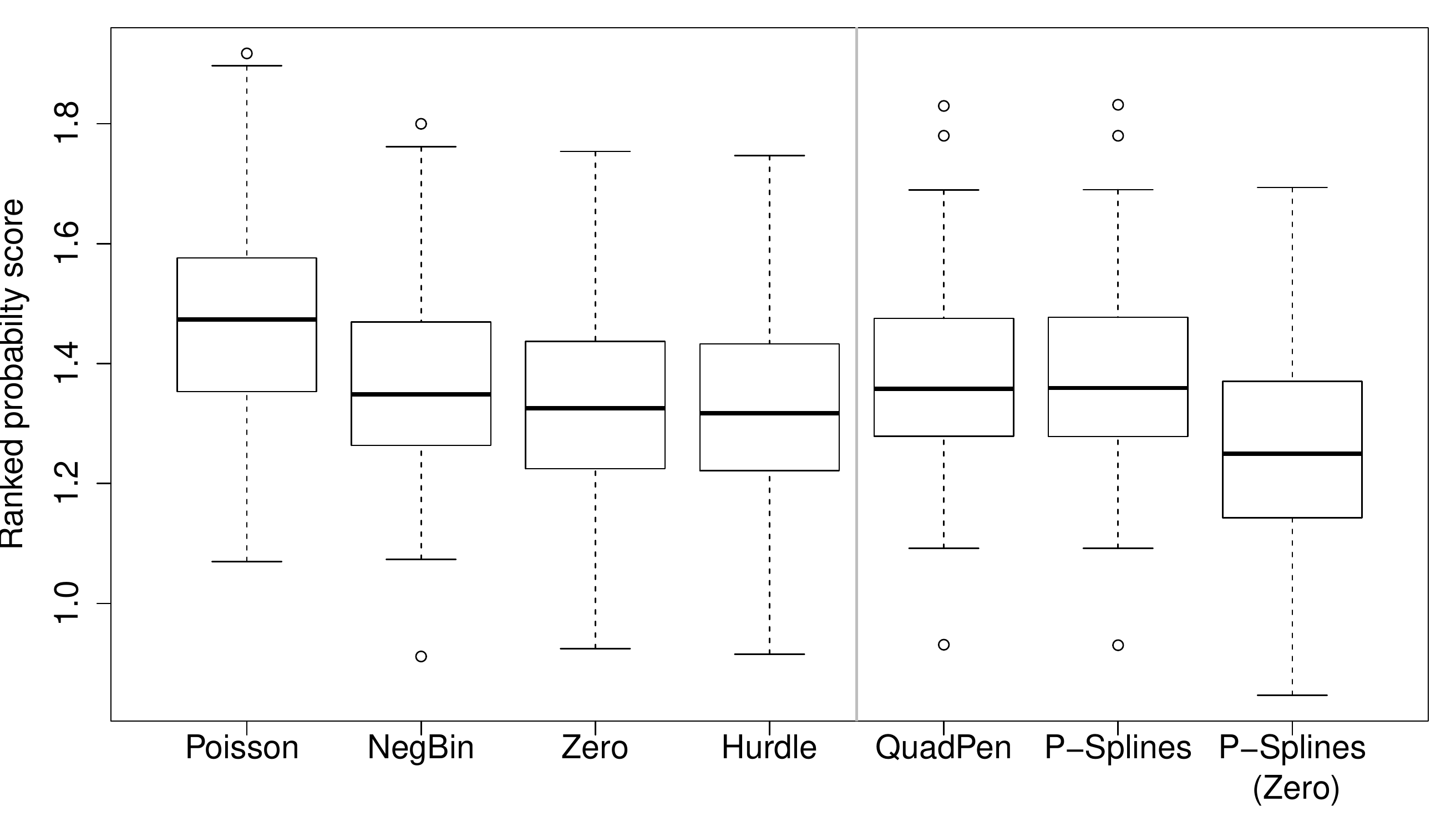}
\caption{Analysis of the boating trips data. The boxplots show the ranked probability score (for $r \in \{0,\hdots,30\}$) of the four classical models (left) and the three transition models (right). All methods were fitted to 100 subsamples without replacement of size $438$ and evaluated on the remaining $219$ observations each.}
\label{fig:rps_rec}
\end{figure}

Next to the approaches (i) - (iv), we also considered the extended transition model \eqref{eq:specific}, referred to as \textit{P-Splines (Zero)}, for model comparison. Figure \ref{fig:rps_rec} shows the ranked probability scores computed from the test data sets (containing 219 observations each) for outcome values $r \in \{0,\hdots, 30\}$. Among the classical models (right panel) the zero-inflation and hurdle model (third and fourth boxplot) showed a better prediction accuracy than the simple Poisson model and the negative binomial model. They also outperformed the basic transition models. The best-performing model (on average), however, was the extended transition model (seventh boxplot), which demonstrates the added value of accounting for excess zeros. 

The results from fitting the extended transition model to the whole sample of $n=657$ individuals is given in Table \ref{tab:coef_rec_zero}. Due to a quasi-complete separation of the outcome with regard to Userfee (all individuals paying a user fee had $>0$~counts), the corresponding coefficient estimate tends to $+\infty$ in the zero part of the model. Also a high quality ranking increases the probability for outcome values greater than zero ($\hat{\beta}_0=1.480$). Within the second part of the model the expected number of boating trips was significantly higher for individuals that (i) gave a positive quality evaluation, (ii) did water-skiing at the lake,  and (iii) payed a user fee. On the other hand, the expected number of boating trips decreased with the expenditure spent when visiting the lake. The continuation ratio \eqref{eq:logmodel} is decreased by the factor $\exp(-1)=0.368$ with an increased expenditure of 100~USD. 

\begin{table}[!t]
\centering
\begin{tabularsmall}{lrrrcrrr}
  \toprule
	&\multicolumn{3}{c}{\bf{Zero}}&&\multicolumn{3}{c}{\bf{Non-Zero}}\\
 & coef & se & z-value && coef & se & z-value \\ 
  \midrule
\bf{Quality}& 1.480 & 0.101 & 14.680 && 0.128 & 0.061 & 2.111 \\ 
\bf{Ski} & 0.244 & 0.318 & 0.765 && 0.454 & 0.156 & 2.912 \\ 
\bf{Income} & -0.031 & 0.083 & -0.379 && -0.085 & 0.053 & -1.621 \\ 
\bf{Userfee} & 13.866 & --- & --- && 1.032 & 0.301 & 3.425 \\ 
\bf{Cost} & -0.003 & 0.003 & -1.066 && -0.010 & 0.002 & -4.687 \\ 
   \bottomrule
\end{tabularsmall}
\caption{Analysis of the boating trips data. Parameter estimates, standard errors and z-values obtained from fitting the extended transition with penalized B-splines accounting for excess zeros.} 
\label{tab:coef_rec_zero}
\end{table}

\section{Software} \label{sec:software}
A crucial advantage of the proposed transition models is that they can be fitted using standard software for binary response models. Before fitting models one has to generate the binary data 
$(\tilde Y_{i0}, \tilde Y_{i1},\dots, \tilde Y_{i,Y_i})^T=(1,1,\hdots,1,0)$ that encode the transitions up to the observed outcome. This is done by the generation of an \textit{augmented data matrix}, which is composed of a set of smaller (augmented) data matrices for each individual. The resulting matrix has $\sum_{i=1}^{n}{Y_i}$ rows. In~R the augmented data matrix can be generated using the function \texttt{dataLong()} in the R~package \textbf{discSurv} \citep{Welchov2019}. Estimates of the model with a quadratic penalty on the intercepts can be computed using the function \texttt{ordSmooth()} in the R~package \textbf{ordPens} \citep{ordPens2015}, estimates of the model with penalized B-splines can be obtained using the function \texttt{gam()} in the R~package \textbf{mgcv} \citep{Wood:2006c}.

\section{Concluding Remarks} \label{sec:conc} 

A semiparametric alternative for the modeling of count data is proposed. The models are very flexible, as they do not assume a fixed distribution for the response variable, but adapt the distribution to the data by using smoothly varying coefficients. The extended form of the model further allows that also the regression coefficients vary smoothly over categories. Importantly, the models directly account for the presence of excess zeros, which has the advantage that no specific two-component model needs to be specified. 

Our applications showed that in terms of prediction (measured by the ranked probability score) the transition models outperform the simple Poisson model and the models tailored to excess zeros, and perform at least as good as the negative binomial model. It was also illustrated that the extension to varying regression coefficients further enhances the flexibility and the predictive ability of the model class. 

An important advantage of the transition models is that they can be embedded into the class of binary regression models. Therefore all inference techniques including asymptotic results to obtain confidence intervals that have been shown to hold for this class of models can be used. A further consequence is that selection of covariates can be done within that framework. Selection of covariates may be demanding even for a moderate number of covariates, because the number of regression coefficients highly depends on the number of response categories. For example, regularization methods as the lasso are applicable but beyond the scope of this article.

We restricted our consideration to the logistic link function. Although it is an attractive choice, because of the simple interpretation of effects, it should be noted that also alternative link function (e.g., the complementary log-log link function) can be used for fitting. Also the assumption of linear predictors can be relaxed by additive predictors, for example, by the use of spline functions for continuous covariates. Then one might investigate non-linear (smoothly varying) coefficients.

\bibliographystyle{chicago}
\bibliography{literatur}
\end{document}